\documentclass{article}
\usepackage{spconf,amsmath,graphicx}
\usepackage{color}

\usepackage{amsfonts,amstext,amssymb,amsthm}
\usepackage{multirow}
\usepackage{subfigure}
\usepackage{setspace}
\usepackage[utf8]{inputenc} % allow utf-8 input
\usepackage[T1]{fontenc}    % use 8-bit T1 fonts
\usepackage{hyperref}       % hyperlinks
\usepackage{url}            % simple URL typesetting
\usepackage{booktabs}       % professional-quality tables
\usepackage{amsfonts}       % blackboard math symbols
\usepackage{nicefrac}       % compact symbols for 1/2, etc.
\usepackage{microtype}      % microtypography
\usepackage{lipsum}
\usepackage{graphicx}
\usepackage{wrapfig}
\graphicspath{ {./images/} }
\usepackage{dsfont}
\usepackage{algorithm,algorithmic}
\usepackage{footmisc}
\def\mycolumn{2}

\usepackage[dvipsnames]{xcolor}

\def\T{{ \mathrm{\scriptscriptstyle T} }}
\newcommand{\norm}[1]{\|#1\|}

\newcommand{\cX}{\mathcal{X}}

\newcommand{\PP}{\mathbb{P}}
\newcommand{\QQ}{\mathbb{Q}}
\newcommand{\RR}{\mathbb{R}}

\newtheoremstyle{mytheoremstyle} % name
    {\topsep}                    % Space above
    {\topsep}                    % Space below
    {\normalfont}                   % Body font
    {}                           % Indent amount
    {\bfseries}                   % Theorem head font
    {.}                          % Punctuation after theorem head
    {.5em}                       % Space after theorem head
    {}  % Theorem head spec (can be left empty, meaning ‘normal’)

\theoremstyle{mytheoremstyle}

\ifx\BlackBox\undefined
\newcommand{\BlackBox}{\rule{1.5ex}{1.5ex}}  % end of proof
\fi

\ifx\QED\undefined
\def\QED{~\rule[-1pt]{5pt}{5pt}\par\medskip}
\fi

\ifx\proof\undefined
\newenvironment{proof}{\par\noindent{\bf Proof\ }}{\hfill\BlackBox\\[2mm]}
\fi

\ifx\theorem\undefined
\newtheorem{theorem}{Theorem}
\fi
\ifx\example\undefined
\newtheorem{example}{Example}
\fi
\ifx\property\undefined
\newtheorem{property}{Property}
\fi
\ifx\lemma\undefined
\newtheorem{lemma}{Lemma}
\fi
\ifx\proposition\undefined
\newtheorem{proposition}{Proposition}
\fi
\ifx\remark\undefined
\newtheorem{remark}{Remark}
\fi
\ifx\corollary\undefined
\newtheorem{corollary}{Corollary}
\fi
\ifx\definition\undefined
\newtheorem{definition}{Definition}
\fi
\ifx\conjecture\undefined
\newtheorem{conjecture}{Conjecture}
\fi
\ifx\fact\undefined
\newtheorem{fact}{Fact}
\fi
\ifx\claim\undefined
\newtheorem{claim}{Claim}
\fi
\ifx\assumption\undefined
\newtheorem{assumption}{Assumption}
\fi
\ifx\condition\undefined
\newtheorem{condition}{Condition}
\fi

\usepackage{cite}

\title{Optimal Sub-sampling \\ to Boost Power of Kernel Sequential Change-point Detection}
\address{School of Industrial and Systems Engineering, Georgia Institute of Technology}
\name{Song Wei \quad Chaofan Huang %\thanks{This work is partially supported by an NSF CAREER CCF-1650913, and NSF DMS-2134037, CMMI-2015787, DMS-1938106, and DMS-1830210.}}
}
\address{School of Industrial and Systems Engineering, Georgia Institute of Technology}
\begin{document}
%\ninept
%
\maketitle
\begin{abstract}
% Sequential change-point detection, which detects distributional shift in the data stream, is a classic and fundamental problem in statistics and machine learning. 
We present a novel scheme to boost detection power for kernel maximum mean discrepancy based sequential change-point detection procedures. Our proposed scheme features an optimal sub-sampling of the history data before the detection procedure, in order to tackle the power loss incurred by the random sub-sample from the enormous history data. We apply our proposed scheme to both Scan $B$ and Kernel Cumulative Sum (CUSUM) procedures, and improved performance is observed from extensive numerical experiments.
\end{abstract}

\section{Introduction}
\label{sec:intro}

Sequential change-point detection is a classic and fundamental problem in signal processing, information theory and statistics, and recently it is of great interest to machine learning and data mining community. The goal is to raise an alarm as soon as possible once a change in distribution occurs in streaming data, and meanwhile to keep the frequency of false alarm as low as possible. 
Classic parametric methods typically handle changes in distribution parameters, but their success heavily relies on exact specification of the probability distribution. As a result, the distribution-free non-parametric methods are much more favorable nowadays.

Kernel maximum mean discrepancy (MMD) \cite{gretton2006kernel,smola2007hilbert,gretton2008kernel,gretton2012kernel} is one of the most popular and powerful non-parametric methods for testing if the two samples are from the same distribution. This can be naturally applied to the change-point detection problem by viewing the history observations and the streaming data as the two samples considered in the test \cite{huang2014high}. However, the quadratic complexity of computing the kernel MMD is the key limitation for its practical use in the online detection setting. Existing methods to tackle such challenge include \cite{li2019scan}, who proposed Scan $B$-procedure by leveraging $B$-test statistic \cite{zaremba2013b} to achieve linear complexity, and \cite{flynn2019change}, who proposed a Kernel Cumulative Sum (KCUSUM) procedure by replacing the likelihood ratio in classic CUSUM with the linear-time MMD statistic. Both methods used a random sub-sample instead of the complete history observations to maintain the computation and memory efficiency. However, such random sub-sample leads to a power loss due to its poor representation of the pre-change distribution. 

Instead, one would want to identify the optimal sub-samples that are most representative of the pre-change distribution, i.e., big data, to minimize the information loss. Kernel herding \cite{chen2010herding} and kernel thinning \cite{dwivedi2021thinning} are two optimal sub-sampling methods based on the kernel MMD statistics, i.e., the sub-samples are chosen such that the kernel MMD between the sub-samples and the big data is minimized. With minimal information loss in the sub-sample selection, one should expect an improvement in the detection power over the use of random sub-samples. 

In this work, we propose a scheme where we perform optimal sub-sampling, or rather, kernel thinning, on history observations as a pre-process before the detection procedure, in order to tackle the poor representation of the random sub-samples. We apply our proposed scheme on both Scan $B$ and KCUSUM procedures. Extensive numerical experiments are conducted to demonstrate that our proposed scheme can help boost the detection power for kernel MMD-based sequential change-point detection procedure.

\section{SETUP}\label{sec:formulation}

Consider sufficiently large amount of history data following pre-change distribution $\PP$ on domain $\cX$ (typically $\cX = \RR^d$):
$$X_1,\dots,X_M \sim \PP, \quad i.i.d.,$$ 
and streaming data $Y_1,\dots,Y_t$ up to current time $t$. In online-change point detection, the goal is to test 
\begin{align*}
    &H_0: X_1,\dots,X_M,Y_1,\dots,Y_t \sim \PP,\\
    &H_1: \exists \  r<t, \  X_1,\dots,X_M,Y_1,\dots,Y_r \sim \PP, \\
    &\quad\quad\quad\quad\quad\quad Y_{r+1},\dots,Y_t \sim \QQ,
\end{align*}
for some unknown distribution $\QQ \not= \PP$.

In sequential change-point detection, we aim to detect the change-point as soon as possible subject to the false alarm rate (i.e., type I error).
The surrogate of false alarm rate is average run length (ARL). 
Typically, we raise an alarm once the detection statistic exceeds a pre-selected threshold $b$, which is properly chosen to satisfy the target ARL requirement. 
We study the expected detection delay (EDD) under such $b$ to investigate the ``power'' of the detection procedure.
For the same target ARL, the smaller the EDD is, the more sensitive/better the detection procedure is. 

\vspace{0.1in}
\noindent
\textit{Notations.} 
We use the subscript to denote the time when the change occurs. To be precise, for a detection procedure defined by a stopping time $T$, we use $\mathbf{E}_{\infty}$ to denote the expectation when all samples follow pre-change distribution $\PP$, and the ARL of such procedure is then defined as $\mathbf{E}_{\infty}[T]$. Similarly, we use $\mathbf{E}_{r}$ denote the expectation when a change occurs at time step $r$. However, it is convenient to consider the case where the change occurs at time step 0, i.e. all streaming data (instead of history observations) come from post-change distribution $\QQ$. Under such case, the EDD is defined as $\mathbf{E}_{0}[T]$. In addition, we denote $a \wedge b = \min\{a,b\}$ and $a \vee b = \max\{a,b\}$.

\subsection{Kernel maximum mean discrepancy statistic}

Before we formally introduce existing kernel MMD-based detection procedure, let us first formally define the kernel MMD statistic.

Given two sets of i.i.d. samples in $\cX$: $\mathbf X = (X_1,\dots,X_m)$ and $\mathbf Y = (Y_1,\dots,Y_n)$, an unbiased estimator of Maximum Mean Discrepancy (MMD) is given by using U-statistic as follows \cite{gretton2012kernel}:
\begin{align}
&\text{\rm MMD}_u^2(\mathbf X, \mathbf Y)= \frac{1}{n(n-1)} \sum_{i=1}^{n} \sum_{j \neq i}^{n} k\left(X_{i}, X_{j}\right) \nonumber \\
&\quad +\frac{1}{m(m-1)} \sum_{i=1}^{m} \sum_{j \neq i}^{m} k\left(Y_{i}, Y_{j}\right) -\frac{2}{m n} \sum_{i=1}^{n} \sum_{j=1}^{m} k\left(Y_{i}, Y_{j}\right) \nonumber.
\end{align}
Under a special case where $m=n$, we can obtain a simpler unbiased estimator:
\begin{align*}
    \text{\rm MMD}_s^2(\mathbf X, \mathbf Y)= \frac{1}{n(n-1)} \sum_{i=1}^{n} \sum_{j \neq i}^{n} h&(X_{i}, X_{j}, Y_{i}, Y_{j}),
\end{align*}
where
\if\mycolumn1
 \begin{equation}\label{eq:kernel_h}
    h(X_{i}, X_{j}, Y_{i}, Y_{j})=k\left(X_{i}, X_{j}\right)+k\left(Y_{i}, Y_{j}\right)-k\left(X_{i}, Y_{j}\right)-k\left(X_{j}, Y_{i}\right),
 \end{equation}
\else
\begin{align}
        h&(X_{i}, X_{j}, Y_{i}, Y_{j}) \label{eq:kernel_h}\\
    &=k\left(X_{i}, X_{j}\right)+k\left(Y_{i}, Y_{j}\right)-k\left(X_{i}, Y_{j}\right)-k\left(X_{j}, Y_{i}\right), \nonumber
\end{align}
%  \begin{equation}\label{eq:kernel_h}
%  \begin{split}
%     h&(X_{i}, X_{j}, Y_{i}, Y_{j})\\
%     &=k\left(X_{i}, X_{j}\right)+k\left(Y_{i}, Y_{j}\right)-k\left(X_{i}, Y_{j}\right)-k\left(X_{j}, Y_{i}\right),
%  \end{split}
%  \end{equation}
\fi
and $k(\cdot,\cdot)$ is the user-specified kernel function:
\begin{equation*}
    k(\cdot,\cdot): \cX \times \cX \rightarrow \RR.
\end{equation*}
Popular kernel function choices include Laplace kernel $k(x,y) = \exp \{-{\norm{x-y}}/{\gamma}\},$ and Gaussian radial basis function (RBF) kernel $k(x,y) = \exp \{-{\norm{x-y}^2}/{\gamma^2}\},$ where $\gamma$ is bandwidth parameter. %Nevertheless, as one will see later in next section, our theoretical analysis is general and allows very flexible kernel choices.

% \begin{figure}[!htp]
% %\vspace{-0.1in}
% \centerline{
% \includegraphics[width = .5\textwidth]{fig/illus/fig0_illus_kerneltrick.pdf}
% }
% %\vspace{-0.1in}
% \caption{Graphical illustration on (i) how to leverage $B$-test statistic \cite{zaremba2013b} to calculate the Scan $B$-statistic \eqref{eq:scan-b} and (ii) the recursive way to calculate our proposed statistic \eqref{eq:proposed_stat} in lines $9 - 14$ in Algorithm~\ref{algo:kernel_CUSUM}.}
% \label{fig:kernel_trick_illus}
% \end{figure}

%\todo{some details on kernel two sample test? linear and B test stat illus}

\subsection{Existing MMD-based detection procedures}

A key drawback of kernel MMD statistic comes from its quadratic computational cost --- as we mentioned earlier, computation efficiency is important in online setting since we need to evaluate the detection statistic in real-time as we receive the streaming data. 

\vspace{.1in}
\noindent
{\it Scan $B$-procedure.} In \cite{li2019scan}, they proposed to use the $B$-test statistic \cite{zaremba2013b} to achieve linear complexity. The first step is radnomly sampling $N \times B$ pre-change samples from the history data $X_1,\dots,X_M$, to form $N$ pre-change blocks with equal block size $B$: $\mathbf{X}_B^{(i)}, \ i = 1,\dots,N$. At time step $t$, the post-change block is constructed by the most recent $B$ streaming observations, i.e., $\mathbf{Y}_B(t) = (Y_{t-B+1},\dots, Y_t)$ and the Scan $B$-statistic is defined as follows:
% As mentioned above, in sequential change-point detection, we need to calculate the detection statistic as we receive the online data. However, MMD \eqref{eq:MMD_estimator} suffers from  due to its .  leveraged  to develop the following Scan $B$-statistic:
\begin{equation*}
    {Z}_t = \frac{1}{N}\sum_{i=1}^N \frac{\text{\rm MMD}_s^2(\mathbf{X}_B^{(i)},\mathbf{Y}_B(t))}{\sigma(B)},
\end{equation*}
By reusing post-change block $\mathbf{Y}_B(t)$ to evaluate the $B$-test statistic, Scan $B$-statistic achieves overall linear computational complexity with respect to (w.r.t.) both time $t$ and (pre-change) sample size $N$, which is much more favorable in online setting. For a pre-selected detection threshold $b$, the Scan $B$ detection procedure is defined by the following stopping time
\begin{equation*}
    T_{B} = \inf \{t : {Z}_t > b\}.
\end{equation*}

The normalizing constant $\sigma(B)$ is the standard deviation of unbiased MMD estimator $\text{\rm MMD}_u^2(\mathbf{X}_B^{(i)},\mathbf{Y}_B(t))$, i.e., 
$$\sigma(B) = \sqrt{\operatorname{Var}\left(\text{\rm MMD}_s^2(\mathbf{X}_B^{(1)},\mathbf{Y}_B(t))\right)}.$$ 
This quantity can be evaluated in closed-form as follows:
\begin{lemma}[Lemma 3.1 \cite{li2019scan}]\label{lma:var_H0}
For block size $B \geq 2$ and the number of pre-change blocks $N > 0$, under $H_0$, we have
\begin{align*}
    \sigma^2(B)=\left(\begin{array}{c}
B \\
2
\end{array}\right)^{-1}\left(C_2 + \frac{C_1 - C_2}{N} \right),
\end{align*}
where $C_1$ and $C_2$ are constants that can be estimated from the history data:
\begin{align*}
\begin{split}
 C_1 & = \mathbf{E}\left[h^{2}\left(X, X^{\prime}, Y, Y^{\prime}\right)\right], \\
 C_2 & = \operatorname{Cov}\left[h(X,X',Y,Y'), h\left(X^{\prime \prime}, X^{\prime \prime \prime}, Y, Y^{\prime}\right)\right].
\end{split}
\end{align*}
Here $X, X^{\prime}, X^{\prime \prime}, X^{\prime \prime \prime}, Y, Y^{\prime} $ are i.i.d. random variables following the pre-change distribution $\PP$.
\end{lemma}

% \subsubsection{KCUSUM}
\vspace{.1in}
\noindent
{\it KCUSUM procedure.}  
An alternative approach leverages linear-time MMD statistic to achieve linear complexity. To be precise, \cite{flynn2019change} replaced the GLR statistic in classic Cumulative Sum (CUSUM) procedure \cite{page1954continuous} with linear-time MMD statistic and proposed the Kernel CUSUM (KCUSUM) procedure. At time $t=0$, the KCUSUM detection statistic is initialized as $S(0) = 0$. 
% To be precise, the linear-time MMD is defined as follows:
% \begin{equation*}
%     \text{\rm MMD}_\ell^2(\mathbf X, \mathbf Y)= \frac{1}{n} \sum_{i=1}^{\lfloor n/2 \rfloor}  h\left(x_{2i-1}, x_{2i}, y_{2i-}, y_{2i}\right),
% \end{equation*}
% where function $h(\cdot,\cdot,\cdot,\cdot)$ is defined in \eqref{eq:kernel_h}. 
\if\mycolumn1
The KCUSUM detection statistic enjoys the following recursive update rule:
\begin{align*}
    S_t=\left\{\begin{array}{ll}
\max\left\{0, S_{t-1} + h\left(x_{1,t}, x_{2,t}, y_{t-1}, y_{t}\right) - \delta \right\} & \text{ if $t$ is even}, \\
S_{t-1} & \text{ if $t$ is odd},
\end{array}\right.
\end{align*}
and $S(0) = 0$. 
\else
The KCUSUM statistic enjoys the following recursive update rule: at each time step $t$, we randomly sample $x^{(t)}$ from the history sample pool. If $t$ is even, then we update it as:
\begin{align*}
     S_t =  0 \vee (S_{t-1} + h\left(x^{(t-1)}, x^{(t)}, y_{t-1}, y_{t}\right) - \delta),  
\end{align*}
otherwise, we keep the detection statistic unchanged, i.e., $
    S_t=  S_{t-1}.$
\fi
The function $h(\cdot,\cdot,\cdot,\cdot)$ is defined in \eqref{eq:kernel_h} and can be viewed as a linear-time MMD statistic with sample size $n = 2$. The hyperparameter $\delta > 0$ makes sure a negative drift.

Similarly, for a pre-selected detection threshold $b$, the KCUSUM detection procedure is defined via the following stopping time
\begin{equation*}
    T_{\rm KCUSUM} = \inf \{t : S_t > b\}.
\end{equation*}

\section{Proposed Method}
Let $\mathbf X = (X_{1},\ldots,X_{M})$ denote the history/pre-change samples. In our proposed procedure, we first compute the optimal sub-samples $\tilde{\mathbf X}=(\tilde x_{1},\ldots,\tilde x_{m})$ with $m < M$ that minimizes over the kernel MMD w.r.t. the pre-change samples, i.e., 
\begin{equation}
    \label{eq:optimal_subsampling}
   \tilde{\mathbf  X} = \arg\min_{\tilde X} \text{\rm MMD}_u^2(\tilde{\mathbf  X}, \mathbf X), \quad \text{ subject to } \tilde x_{i}\in X\; \forall i.
\end{equation}
Herding \cite{chen2010herding} is one greedy approach to solve \eqref{eq:optimal_subsampling} by adding one point at a time to the sub-samples $\mathbf X$. Kernel thinning \cite{dwivedi2021thinning} is a recent proposed approach that also aims to solve \eqref{eq:optimal_subsampling} but with stronger theoretical guarantee. We use kernel thinning in our procedure due to its public available implementation. In our proposed detection procedure, we replace the raw history pre-change samples with the kernel thinned optimal sub-samples and the rest of the detection procedure remains the same. We graphically illustrate our proposed method on Scan $B$ procedure in Figure~\ref{fig:illus} below.

\begin{figure}[!htp]
%\vspace{-0.1in}
\centerline{
\includegraphics[width = .5\textwidth]{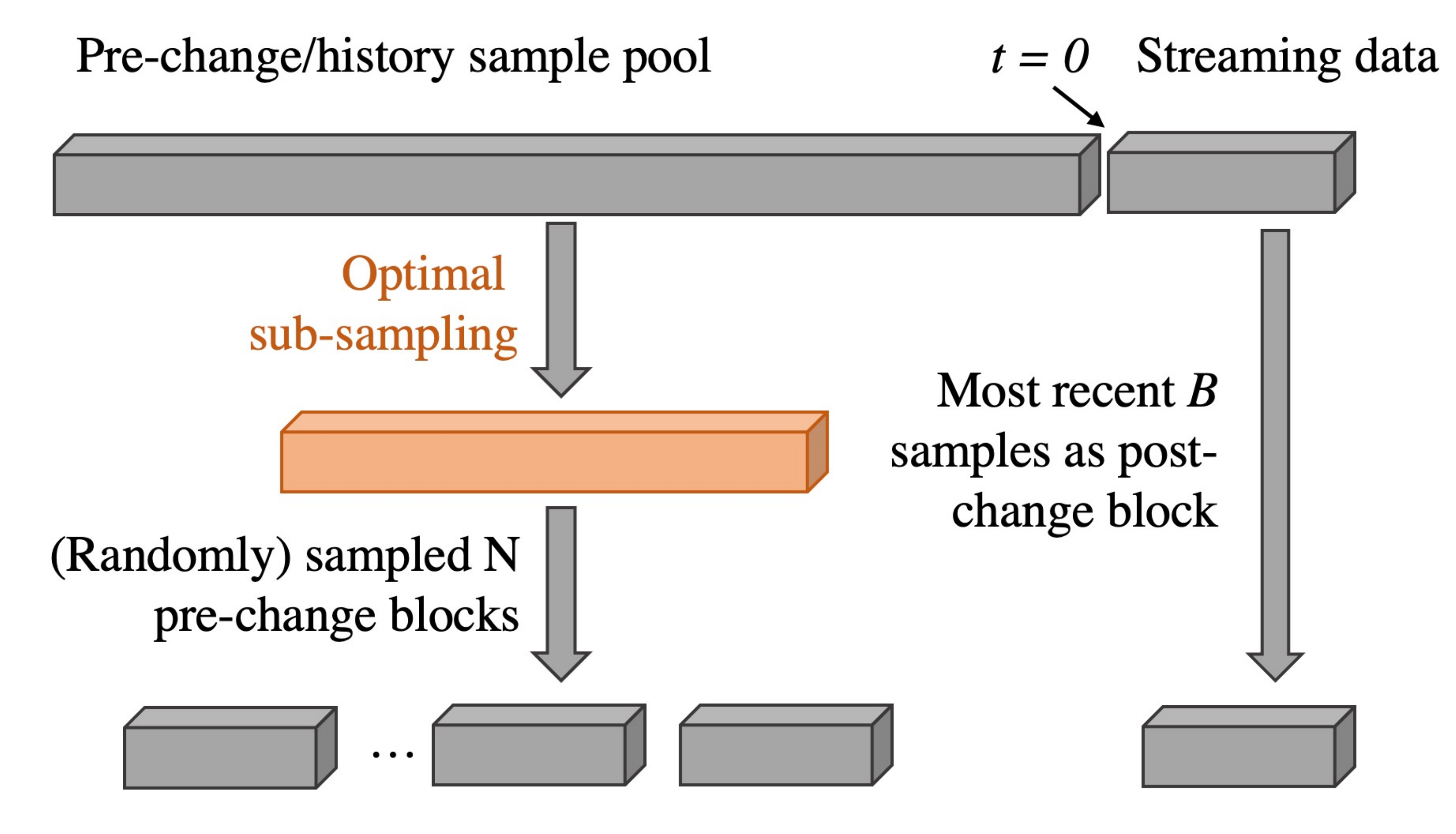}
}
\vspace{-0.05in}
\caption{Illustration of our proposed method: Scan $B$ procedure coupled with optimal sub-sampling.}
\label{fig:illus}
\end{figure}

The motivation of our proposed method comes from the random sampling from the pre-change samples for both Scan $B$ and KCUSUM procedures --- for example, in Scan $B$ procedure, when we want to randomly draw a small portion of data to represent the pre-change distribution $\PP$, it would be much easier to sample from kernel thinned pre-change samples than the enormous raw pre-change samples. By performing such optimal sub-sampling, the variance of the random sampling in both procedures will be reduced, which will result in better detection power. 

\section{Numerical examples} 

\begin{figure*}[!htp]
%\vspace{-0.1in}
\centerline{
\includegraphics[width = 0.9\textwidth]{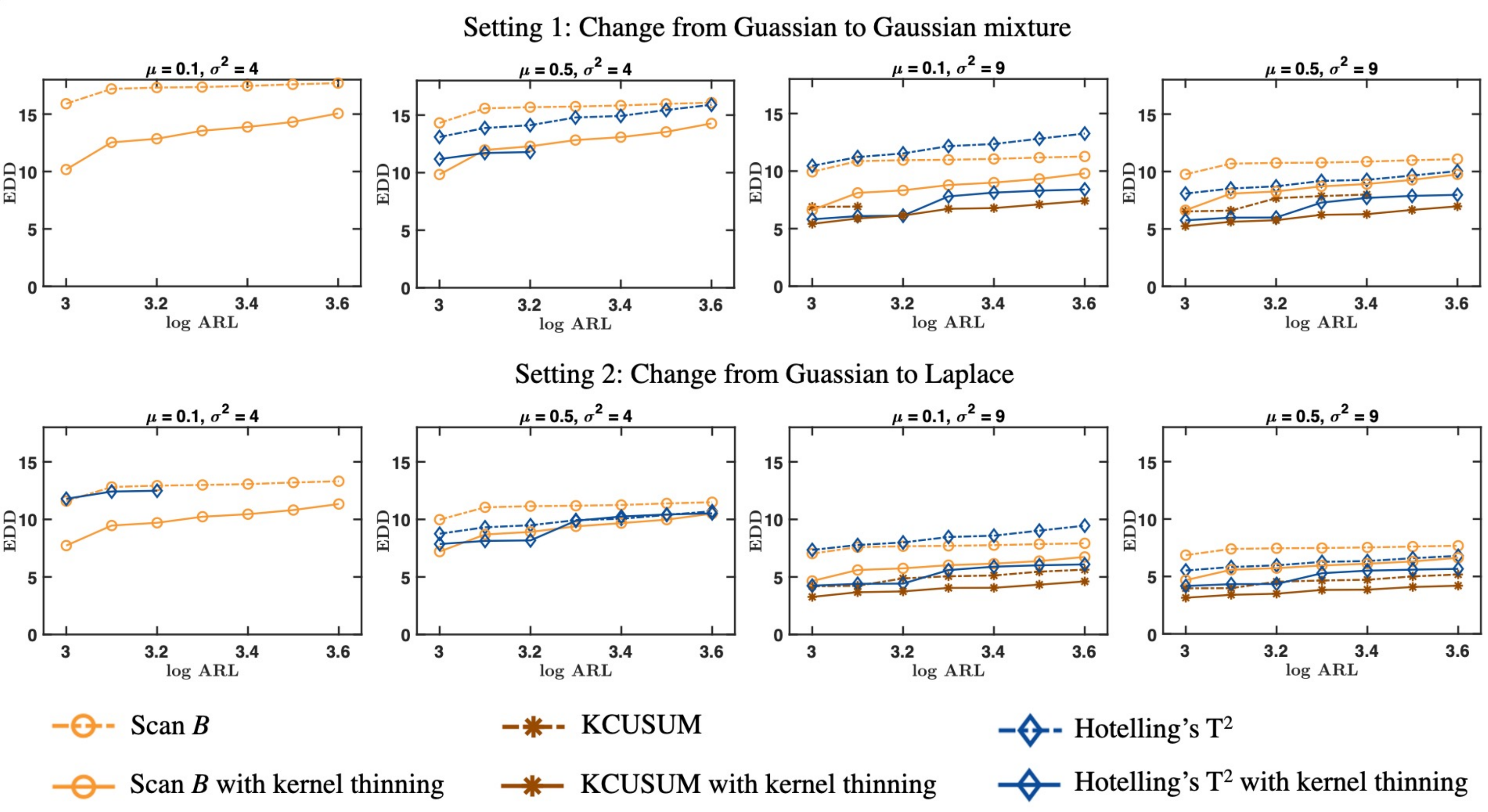}
}
\vspace{-0.1in}
\caption{Expected detection delay (EDD) for given average run length (ARL). The absence of the dot indicates the corresponding method fails to detect the change before time $t = 50$ under the corresponding setting. We can observe that both Scan $B$ and KCUSUM procedures (even Hotelling's $T^2$ procedure) achieve better detection power, i.e., smaller EDD, after we perform kernel thinning on pre-change samples as a pre-process.}
\label{fig:EDDARL_compare_comprenhensive}
\vspace{-0.1in}
\end{figure*}

In this section, we numerically verify the good performance of our proposed method. We will compare the detection power with and without optimal sub-sampling on the history sample pool. In addition to aforementioned Scan $B$ and KCUSUM detection procedures, we also consider a parametric benchmark method, which is Hotelling's $T^2$ procedure.

\vspace{.1in}
\noindent
{\it Hotelling's $T^2$ procedure.}  As one of the most classic and commonly used parametric two sample test statistic, Hotelling's $T^2$ statistic can be naturally adapted to sequential change-point detection. To be precise, at time step $t$, for hypothetical change-point $r < t$, we split the data into two parts:
\begin{align*}
    U &= (X_1,X_2,\dots,X_M,Y_1,\dots,Y_{r-1}), \\
    V &= (Y_{r},\dots,Y_t).
\end{align*}
For notational simplicity, we denote the elements in $U$ and $V$ by $U_1,\dots,U_{M + r - 1}$ and $V_1,\dots,V_{t - r + 1}$, respectively. 
We define
$$
HT^{2}_t(r)=\frac{(M + r - 1)(t - r + 1)}{M + t}\left(\bar{U}-\bar{V}\right)^{\T} \widehat{\Sigma}^{-1}\left(\bar{U}-\bar{V}\right),
$$
where superscript $^\T$ standards for vector/matrix transpose and
\begin{align*}
    \bar{U} = \sum_{i=1}^{M+r-1} U_i/(M + r - 1), \ \bar{V} =  \sum_{i=1}^{t-r+1} V_i/(t - r + 1).
\end{align*}
$\widehat{\Sigma}$ is the pooled covariance matrix and estimated as follows:
\if\mycolumn1
\begin{align*}
    \widehat{\Sigma}=(M + t-2)^{-1}\Bigg(&\sum_{i=1}^{M+r-1}\left(U_{i}-\bar{U}\right)\left(U_{i}-\bar{U}\right)^{\T} +\sum_{i=1}^{t-r+1}\left(V_{i}-\bar{V}\right)\left(V_{i}-\bar{V}\right)^{\T}\Bigg).
\end{align*}
\else
\begin{align*}
    \widehat{\Sigma}=(M + t-2)^{-1}\Bigg(&\sum_{i=1}^{M+r-1}\left(U_{i}-\bar{U}\right)\left(U_{i}-\bar{U}\right)^{\T} \\
    &+\sum_{i=1}^{t-r+1}\left(V_{i}-\bar{V}\right)\left(V_{i}-\bar{V}\right)^{\T}\Bigg).
\end{align*}
\fi
The Hotelling's $T^2$ detection statistic is defined as follows:
\begin{equation*}%\label{eq:HT2}
    S_t^{\rm H} = \max_{1\leq r \leq t-1} HT^{2}_t(r).
\end{equation*}   
And the corresponding detection procedure is defined as:
\begin{equation*}
    T_{{\rm HT}^2} = \inf \{t : S_t^{\rm H} > b\},
\end{equation*}
where $b$ is a pre-selected detection threshold.

\vspace{0.1in}
\noindent \textit{Experimental settings.}
For Scan $B$ and KCUSUM detection procedures as well as the optimal sub-sampling via MMD (i.e., kernel thinning), we all choose Gaussian RBF kernel function, whose the bandwidth parameter is chosen to be the median of the $\ell_2$ distance matrix for pre-change samples (this is known as median heuristic and performs well in practice \cite{ramdas2015decreasing}).
For Scan $B$ procedure, we choose $N=15$ pre-change blocks with block size $B = 50$, as suggested by the numerical evidence in \cite{li2019scan}. 
For KCUSUM procedure, we choose $\delta = 1/50$ as suggested by the original paper \cite{flynn2019change}. 

We consider the change from $20$-dimensional standard Gaussian distribution $N(\mathbf{0}, I_{20})$ to: 
\begin{itemize}
    \item [1.] Gaussian mixture: $N(\mathbf{0}, I_{20})$ with probability (w.p.) 0.3; $N(\mu \mathbf{1},\sigma^2 \ I_{20})$ w.p. 0.7;
    \item [2.] $20$-dimensional i.i.d. Laplace distribution Lap$(\mu,\sigma)$.
\end{itemize}
By kernel thinning \eqref{eq:optimal_subsampling}, we obtain a size-$2,500$ optimal sub-sample out of $10,000$ raw pre-change samples.

The detection threshold $b$ is chosen separately for each detection procedure via Monte Carlo simulation, in order to satisfy the corresponding ARL requirement. We aim to numerically compare the EDD for given ARL.

\vspace{0.1in}
\noindent \textit{Results.} As we mentioned earlier, for a given false alarm rate (i.e., ARL), the smaller the EDD is, the better the procedure will be.
We plot the EDD against logarithm of ARL (base $10$) for those aforementioned methods in Figure~\ref{fig:EDDARL_compare_comprenhensive}.

In Figure~\ref{fig:EDDARL_compare_comprenhensive}, the absence of the dot indicates the corresponding method fails to detect the change before time $t = 50$. We can observe: 
(i) First and foremost, for both Scan $B$ and KCUSUM detection procedures, kernel thinning on the history data can boost the detection power, i.e., smaller EDD for given ARL; moreover, even for Hotelling's $T^2$ procedure, we can see its power is boosted when coupled with kernel thinning in most cases.
(ii) Secondly, the non-parametric Scan $B$ procedure is more robust, i.e., it can still detect the change in all cases considered above, even when the change is small; in contrast, KCUSUM procedure is more sensitive, i.e., it can achieve quicker detection under certain cases.
(iii) Lastly, the Hotelling's $T^2$ procedure's performance is ``in between'': on one hand, compared with KCUSUM procedure, it is a little bit more robust and can detect the small change under some cases; on the other hand, as a parametric approach, it performs better when the change is larger (especially for large mean shift), when compared with Scan $B$ procedure.

\section{Conclusion}

This work developed a scheme to boost detection power for kernel MMD-based sequential change-point detection procedures by leveraging kernel thinned history data. Ongoing work includes (i) directly using the kernel thinned samples as the pre-change blocks (without additional random sub-sampling) (ii) exploring the effect of different optimal sub-sampling methods, such as support points \cite{mak2018sp}.

% \textcolor{blue}{Other than the kernel herding and kernel thinning that are built upon the kernel MMD, \cite{mak2018sp} also recently proposed the support points optimal subsampling method that is built upon the energy distance\cite{szekely2013es}, another popular two-sample test statistics. Different from the kernel MMD, no tuning parameter, e.g., kernel bandwidth, is required for support points. There are also subsampling methods based on the kernel stein discrepancy, but this require knowing the unnormalized target distribution and its gradient, which might not be applicable in the change-point detection setting.}

% \section*{Acknowledgment}

% \section{REFERENCES}
% \label{sec:refs}

% List and number all bibliographical references at the end of the
% paper. The references can be numbered in alphabetic order or in
% order of appearance in the document. When referring to them in
% the text, type the corresponding reference number in square
% brackets as shown at the end of this sentence \cite{C2}. An
% additional final page (the fifth page, in most cases) is
% allowed, but must contain only references to the prior
% literature.


\newpage

\bibliographystyle{IEEEbib}
\bibliography{icasspref}

\end{document}